\documentclass{elsart}

\def\CuHeu{\textrm{\textit{Cu$_{2}$MnAl}}}
\def\CoSnHeu{\textrm{\textit{Co$_{2}$MnSn}}}
\def\CoSiHeu{\textrm{\textit{Co$_{2}$MnSi}}}
\def\CoGeHeu{\textrm{\textit{Co$_{2}$MnGe}}}
\def\CoSnHeu{\textrm{\textit{Co$_{2}$MnSn}}}
\def\CuHeuCoGeHeu{\textrm{[\textit{Cu$_{2}$MnAl$_{(3nm)}$/Co$_{2}$MnGe$_{(3nm)}$}]$_{30}$}}
\def\CuHeuCoSnHeu{\textrm{[\textit{Cu$_{2}$MnAl$_{(3nm)}$/Co$_{2}$MnSn$_{(3nm)}$}]$_{30}$}}
\def\CuHeuAu{\textrm{[\textit{Cu$_{2}$MnAl$_{(3nm)}$/Au$_{(3nm)}$}]$_{30}$}}
\def\CoSnHeuAu{\textrm{[\textit{Co$_{2}$MnSn$_{(3nm)}$/Au$_{(3nm)}$}]$_{30}$}}
\def\CoSnHeuV{\textrm{[\textit{Co$_{2}$MnSn$_{(3nm)}$/V$_{(3nm)}$}]$_{30}$}}
\def\CoGeHeuAu{\textrm{[\textit{Co$_{2}$MnGe$_{(3nm)}$/Au$_{(3nm)}$}]$_{30}$}}
\def\CoGeHeuV{\textrm{[\textit{Co$_{2}$MnGe$_{(3nm)}$/V$_{(3nm)}$}]$_{30}$}}
\def\deg{$^\circ$}

\usepackage{graphicx}
\usepackage{dcolumn}
\usepackage{bm}
\usepackage[bf]{subfigure}

\begin{document}

\begin{frontmatter}

\title{Preparation and structural properties of thin films and multilayers of the Heusler
compounds Cu$_{2}$MnAl, Co$_{2}$MnSn, Co$_{2}$MnSi and
Co$_{2}$MnGe }
\author{U. Geiersbach}
\author{A. Bergmann}
\author{K. Westerholt}
\ead{Kurt.Westerholt@ruhr-uni-bochum.de}
\address{ Institut f\"ur
Experimentalphysik IV, Ruhr-Universit\"at D44780 Bochum, Germany }

\date{\today}
\bigskip
\bigskip

\begin{abstract}

We report on the preparation of thin films and multilayers of the
intermetallic Heusler compound {\CuHeu}, {\CoSnHeu}, {\CoSiHeu} and
{\CoGeHeu} by rf-sputtering on MgO and Al$_{2}$O$_{3}$ substrates.
{\CuHeu} can be grown epitaxially with (100)-orientation  on MgO (100)
and in (110)-orientation on Al$_{2}$O$_{3}$ a-plane.  The Co based
Heusler alloys need metallic seedlayers to induce high quality textured
growth. We also have prepared multilayers with smooth interfaces by
combining the Heusler compounds with Au and V. An analysis of the
ferromagnetic saturation magnetization of the films indicates that the
{\CuHeu}-compound tends to grow in the disordered $B2$-type structure
whereas the Co-based Heusler alloy thin films grow in the ordered $L2_1$
structure. All multilayers with thin layers of the Heusler compounds
exhibit a definitely reduced ferromagnetic magnetization indicating
substantial disorder and intermixing at the interfaces.\\

\end{abstract}
\end{frontmatter}

\clearpage
\section{\label{sec:level1}Introduction\protect\\ }

The new, rapidly evolving field of magnetoelectronics \cite{art:park98}
started an upsurge of interest in ferromagnetic metals with full spin
polarization at the Fermi level. In principal these so called half
metallic ferromagnets are ideal for applications in tunnelling
magnetoresistance (TMR)\cite{art:Moodera} or giant magnetoresistance
(GMR) \cite{art:Dieny} elements and as electrodes for spin polarized
current injection into semiconductors. \\
Half metallic ferromagnetic alloys are scarce, since usually the s- and
p-type valence electrons contribute both spin directions at the Fermi level.
From electronic energy band structure calculations on knows several
ferromagnetic oxides like $CrO_{2}$ \cite{art:Schwarz} and
$La_{1-x}Sr_{x}MnO_{3}$ \cite{art:park98}. Until now there are only a few
intermetallic compounds known to have this unique property, all belonging to
the Heusler group with the general formula $A_2BX$ (A=Cu, Co, Ni…, B=Mn,
Fe...,  X=Al, Ge, Si)\cite{pbk:ziebeck}. The basic ordered Heusler structure
is a cubic lattice (space group Fm3m) with four interpenetrating fcc
sublattices occupied by A, B or X- atoms respectively. There are several
structural variants of the Heusler unit cell with different degrees of site
disorder of the atoms on the A, B and X-positions. Among them the B2
structure with a random occupancy of the B and X-position and the completely
disordered bcc structure with a random occupancy on the A, B and
X-positions \cite{pbk:ziebeck}. \\
The ferromagnetic half metals known from theoretical electron energy band
structure calculations are the compounds $PtMnSb$ and $NiMnSb$
\cite{art:deGroot83} (so called half Heusler compounds since one
A-sublattice is empty) and the compounds {\CoSiHeu} and {\CoGeHeu}
\cite{art:ishida2-98}. {\CoSnHeu} and $Co_2MnSb$ in a strict sense do not
belong to this group, since they possess only about 90 \% of spin
polarization at the Fermi level, but they can be made half metallic
ferromagnets by alloying \cite{art:ishida1-95}.\\
In recent years the properties of thin films of the half Heusler compounds
$PtMnSb$ and $NiMnSb$ have been studied intensely by several groups in order
to elucidate their potential in the field of magnetoelectronics
\cite{art:kabani90}. These compounds have also been tested already in TMR-
and GMR-thin film devices \cite{art:caballero99}, however with only moderate
success until now. The main difficulty one encountered when preparing thin
films of the half Heusler compounds is a high degree of site disorder. This
leads, on the one hand, to strong electron scattering and a low electron
mean free path which has a negative influence of the amplitude of the GMR
\cite{art:Kautzky}. On the other hand, it is expected that the site disorder
destroys the full spin polarization at the Fermi level, which theoretically
has been predicted only when assuming perfectly ordered $A_2BX$ structure
with pure $L2_1$ type of site symmetry \cite{art:Orgassa}. \\

The half metallic ferromagnets from the Heusler group {\CoGeHeu},
{\CoSiHeu}, and $Co_2MnSn_{1-x}Sb_x$ found much less attention in the
experimental literature until now. Recently two groups published first
investigations of GMR elements using {\CoSiHeu} \cite{art:raphael01} and
{\CoGeHeu} \cite{art:ambrose01}. Similar to the results on the spin valves
based on the half Heusler compounds the amplitude of the GMR was found to be
very low, the reason for this was not clear. We have presented our first
experimental results on thin films of the Co-based Heusler compounds in
\cite{art:Geiersbach}. In this paper we present in the first part a detailed
study of the preparation and structural properties of thin films of the
Heusler phases {\CoSiHeu}, {\CoGeHeu}, {\CoSnHeu} and {\CuHeu}. The latter
Heusler phase is a ferromagnet but not half metallic \cite{art:kuebler83}.
We use it as a reference compound and as a seed layer for improving
the growth of the Co-based Heusler alloys. \\

In the second  part of the present paper we report on  multilayers prepared
by combining thin layers of two different Heusler compounds and Heusler
compounds with non magnetic metals. Multilayers with Heusler compounds have
rarely been studied in the literature until now, we only know of
publications on [PtMnSb/NiMnSb] multilayers \cite{art:Mancoff}. Our original
intention to study Heusler-based multilayers was to search for an
antiferromagnetic interlayer exchange coupling (IEC). Quantum interference
models for the IEC suggest that it should exist in virtually any multilayer
system combining ferromagnetic and non ferromagnetic metallic layers
\cite{Bruno}. Experimentally, however, a prerequisite for the observation of
the IEC is a high quality of the layered structure with flat interfaces. Our
results to this end were negative until now. We could not find  clear
evidence for an antiferomagnetic IEC in the multilayer systems we report on
in the next section. Instead, we use the multilayers here mainly as a tool
for gaining insight into the magnetic properties of the Heusler alloys films
in the limit of a very small thickness.

\section{\label{sec:level1}Preparation and Experimental\protect\\ }

Our thin films and multilayers were deposited by rf-sputtering using pure Ar
at a pressure of $5\cdot10^{-3}$ mbar as sputter gas. The base pressure of
the sputtering system was $5\cdot10^{-8}$ mbar, the sputtering rate was 0.04
nm/s for the Heusler compounds, 0.06 nm/s for Au and 0.03 nm/s for V. For
the growth of pure Heusler alloy thin films the temperature of the
substrates was $470^\circ$C, the multilayers were grown at a temperature of
$300^\circ$C. A systematic change of the process parameters
showed that these values gave the best structural results.\\
Heusler alloy targets with 10 cm diameter have been made from single
phase, stoichiometric ingots prepared by high frequency melting of the
components in high purity graphite crucibles. The thin films  of the
present study were grown on a-plane sapphire substrates or MgO
(100)-substrates which were carefully cleaned and ion beam etched
prior to deposition.\\
During the sputter deposition process of the multilayers the substrates were
moved automatically between the two targets of the dual source discharge.
After finishing 30 periods of the multilayers we deposited a 2 nm thick
Au-cap layer at room temperature for protection against oxidation. We
usually prepared series of 10 multilayers simultaneously within the same run
with either the thickness of the Heusler compound or the thickness of the
other metal varied. The thickness covered typically a range between  0.6 nm
and 3 nm for each component. For preparing the constant layer thickness the
substrate holder was rotated in the symmetric position above the target, for
the preparation of the variable thickness we made use of the natural
gradient of the deposition rate when the substrate holder is in an
off-centric position.
\\
The stoichiometric composition of our thin films was controlled using
quantitative electron microprobe analysis. For \CoSiHeu there is a small
Si-deficiency (23~at.\% Si instead of 25~at.\%), probably caused by
selective resputtering of Si, for \CuHeu we find some excess of Cu (52~at.\%
instead 50 at.\%). For the other two Heusler phases of the present study the
thin films preserve the stoichiometric composition of the targets to within
the precision of the microprobe analysis of about 0.5~at.\%.\\
The structural characterization of all samples was carried out by a thin
film 3-circle x-ray spectrometer using $Cu-K_\alpha$-radiation. The x-ray
study combined small angle reflectivity, $\Theta-2\Theta$ Bragg scans and
rocking scans with the scattering vector out of the film plane. For selected
samples Bragg scans and rocking scans at glancing incidence with the
scattering vector in the film plane were also taken. The determination of
the saturation magnetization was performed by a commercial SQUID
magnetometer (Quantum Design MPMS system) at a temperature of 5 K and at a
field of 0.4 T, which is far above the coercive force for all samples under
study here.

\section{\label{sec:level1}Results and discussion\protect\\ }

\subsection{\bf\label{sec:level2}{\CuHeu} thin films}

We first discuss the properties of thin films of the {\CuHeu} Heusler phase.
In the bulk the Fm3m phase of this compound is not stable below 923 K but
decomposes into the phases $\beta-Mn$, $\gamma-Cu_9Al_4$ and $Cu_3AlMn_2$
\cite{art:Soltys}. Interestingly we found, however, that single phase thin
films can be prepared by sputtering on MgO and sapphire a-plane at
$470^\circ$ C and are metallurgically stable even when annealed for a long
time at this temperature. Thus the sputtering process and the epitaxial
strain seems to establish stability conditions definitely different from the
bulk. {\bf Fig.1} shows an x-ray Bragg scan over the whole angular range of
a {\CuHeu} film with a thickness of 100 nm grown on MgO (100). One observes
only the Heusler (200) and (400)-peak indicating perfect epitaxial
(100)-growth. The out-of-plane rocking width of the (200) Bragg peak was
determined to be $0.16^\circ$. In the inset of Fig.1 we present the in-plane
rocking scan of the Heusler (200) reflection exhibiting 4 peaks at a
distance of $90^\circ$, as expected for a single crystalline layer. The
[010] direction of the Heusler film is rotated by an angle of $45^\circ$
from the in-plane MgO-[010]-direction. An example of the growth of the
{\CuHeu}-phase on $Al_2O_3$ a-plane is shown in {\bf Fig.2a}. One observes a
perfect out-of-plane (220)-texture of the Heusler phase with an out-of-plane
rocking width of $0.8^\circ$ for the (220)-Bragg-peak.  An in-plane rocking
scan of the (220)-Bragg peak, however, reveals an in-plane polycrystalline structure.\\
The {\CuHeu} thin films prepared on MgO and on sapphire a-plane at $470
^\circ$ C possess a flat morphology. As an example we show low angle
x-ray reflectivity spectra of the {\CuHeu}-film in {\bf Fig.2b}. One
observes well defined thickness oscillations up to scattering angles of
$2\Theta\approx 5^\circ$. From a simulation of the reflectivity spectrum
using the Parratt formalism \cite{art:paratt54} we derive a total
thickness of 108 nm and estimate a roughness parameter of 0.6 nm. Atomic
force microscopic images of the surfaces also show a very flat surface
morphology with a roughness of about 0.7 nm (rms).\\
An important characterization of the metallurgical state of {\CuHeu} films
is the degree of order between the sites A, B and X of the Heusler unit
cell. In polycrystalline, bulk material the relative intensity of the
superstructure Bragg reflection (111) is conveniently used to determine the
order parameter S for the site order between the B-(Mn) and X-(Al)
positions, S=1 defining perfect order ($L2_1$- structure) and S=0 defining
complete site disorder (B2-structure)\cite{art:Bacon}. Unfortunately for our
epitaxial (100) or (110)-films the (111)-Bragg peak is not accessible by our
triple axis x-ray spectrometer. Qualitatively the degree of site disorder
can be deduced from the value ferromagnetic saturation magnetization
\cite{art:Robinson}. {\CuHeu} single crystals with perfect site order
$S\approx1$ have a saturation magnetization $M_s=98$ emu/g corresponding to
a magnetic moment of about $4.2~\mu_B/Mn-atom$, B2-type disorder leads to a
decrease of $M_{s}$, since Mn-spins on the X-position do not couple
ferromagnetically to the Mn-spins on the B-position. {\CuHeu} in the
completely disordered B2-state exhibits spin glass order with a very low
value of the magnetization \cite{art:Taylor}. For the single crystalline
{\CuHeu} film on MgO we get $M_s=40~emu/g$ pointing towards a substantial
degree of site disorder. For the film prepared on a-plane $Al_2O_3$ we get
$M_s=62~emu/g$, this value comes closer to the bulk value for $M_s$. The
structural parameters and the saturation magnetization for the {\CuHeu}
phase are summarized in {\bf Table 1}. Note that the reduction of the moment
correlates with a definite decrease of the lattice parameter.

\subsection{\bf\label{sec:level2} {\CoSiHeu}, {\CoGeHeu} and {\CoSnHeu} thin films }

The {\CoSiHeu}, {\CoGeHeu} and {\CoSnHeu} halfmetallic Heusler thin films
grown directly on MgO or Al$_2$O$_3$ are polycrystalline and have a bad
structural quality and a low value for the saturation magnetization. Only
when using suitable metallic seed layers with a typical thickness of
about 2 nm we could achieve textured growth and good structural quality.
For the {\CoSnHeu} phase we found that the optimum seed layer for the
growth on sapphire a-plane is Au with a lattice parameter mismatch of
about 1\%. V and {\CuHeu} seedlayers can also be used. In {\bf Fig.3} we
show an out-of-plane Bragg-scan of a {\CoSnHeu}-film grown on an Au seed
layer. One observes only the (220)- and the (440)-Heusler-Bragg-peak,
evidencing pure (110)-texture. The rocking width of the (220) Heusler
Bragg-peak is about 3{\deg} i.e. it is definitely larger than obtained
for the {\CuHeu} layers (see Table 1). The {\CoSiHeu} phase with similar
structural quality can be grown on V and Cr seedlayers, Cr giving a
 slightly better growth quality since it has a lattice mismatch of 0.8\%
only. For the {\CoGeHeu}-phase V, Au and Cr-seed layers give comparably
good structural quality. A summary of these results is given in Table 1.
 For all three Co-based Heusler compounds the thin films have a
very flat surface morphology. As one representative example we show a
small angle x-ray reflectivity scan of the {\CoGeHeu} film grown on a
V-seedlayer in {\bf Fig.4}. One observes well defined finite thickness
oscillations from the total layer superimposed by an oscillation from the
V-seedlayer. From a fit using the Parrat formalism we estimate a
roughness of about 0.5 nm for the interfaces.\ We also have tested
systematically the growth of the Co-based Heusler phases on MgO (100)
substrates. The films grown on the bare MgO-surface are polycrystalline.
When using metallic seedlayers one can induce reasonable quality
out-of-plane (100)-textured growth, however with a definitely larger
mosaicity than for the growth on sapphire a-plane, as evidenced by the
increased rocking width of the Bragg peaks (see Table 1). Contrary to the
case of the {\CuHeu} phase we could not achieve epitaxial growth for the
Co-based Heusler alloys, the structure in-plane is always polycrystalline
with a broad distribution of the (220) or (200)
Bragg peak intensity for an in-plane rocking scan.\\
 \\
An important criterion for the magnetic quality of the thin films is the
value of the ferromagnetic saturation magnetization $M_s$ which we have
included in Table 1. The values for $M_s$ for the Co-based Heusler alloy
thin films are close to the bulk values and for the {\CoSnHeu} and the
{\CoGeHeu}-phase nearly coincide with them. This indicates the absence of
sizable B2-type of site disorder, consistent with the fact that for these
phases the ordered $L2_1$-type phase is very stable \cite{art:Webster}.
Only for the {\CoSiHeu}-thin film we observe a definitely smaller value
of the magnetization in the film than in the bulk, which we would
attribute to the deviation from the ideal stoichiometry
for this film.\\
The standard growth temperature we apply for the growth of the films in
Table~1 was 470 \deg C. In thin film heterostructures the maximum
temperature which is allowed for avoiding strong interdiffusion of the
components at the interfaces is often definitely lower. Thus it is essential
to know the change of site disorder and the sample quality when applying
lower substrate temperatures. We prepared series of films of the Heusler
alloys at lower substrate temperatures down to T=100 \deg C. We found that
the structural quality, as judged from the Bragg reflection intensity, is
only slightly worse when preparing at 300\deg C, at still lower preparation
temperatures, however, there is a definite deterioration of the crystal
quality. Simultaneously the ferromagnetic saturation magnetization is
strongly reduced for the {\CuHeu} phase {\bf(Fig.5)} and moderately for the
{\CoSnHeu} phase and the {\CoGeHeu} phase . This indicates an
increasing degree of site disorder when lowering the preparation temperature.\\

\subsection{\label{sec:level2}Multilayers with Heusler alloys}

In this section we want to elucidate the possibility to grow multilayers
based on the {\CuHeu}, {\CoSnHeu} and the {\CoGeHeu} compounds. In order to
avoid excessive interdiffusion at the interfaces the substrate temperature
during the preparation of the multilaysers had to be limited to 300 \deg C,
although at this temperature the ferromagnetic saturation magnetization of
the Heusler compounds is already definitely reduced (see Fig.5). Actually at
300 \deg C multilayers with high structural quality of all these phases can
be grown on sapphire a-plane by combining them with fcc Au. {\bf{Fig.6a}}
shows a small angle x-ray reflectivity scan of a {\CuHeuAu} multilayer with
a nominal thickness, as calculated from the sputtering rate, of 3 nm for Au
and {\CuHeu} combined of 30 periods. Above the critical angle for total
reflection $\Theta_c$ the multilayer structure gives rise to superlattice
reflections superimposed on the Fresnel-reflectivity. We observe
superlattice reflections up to 4th order, revealing a good interface quality
and low fluctuations of the layer thickness. From the reflectivity peak of
order l at the angle $\Theta_{l}$ one can calculate the superlattice
periodicity $\Lambda$ by using  the relation \cite{art:zabel90}

\begin{eqnarray}
\Lambda&=&l\cdot\lambda/[2(\sqrt{\Theta_{l}^{2}-\Theta_{c}^{2}})]
\end{eqnarray}

From a fit we get $\Lambda=5.7$ nm in good agreement with the nominal
thickness. From simulations of the reflectivity curves using the Parratt
formalism \cite{art:paratt54} we derive an interface roughness of about 0.6
nm. The out-of-plane Bragg scan {\bf(Fig.6b)} close to the (220)/(111)
fundamental Bragg reflection reveals that the multilayer possesses a pure
(110) out-of-plane texture for {\CuHeu}, and (111) texture for the
Au-layers. Besides the fundamental Bragg peak from the average lattice, the
multilayer exhibits a rich satellite structure caused by the chemical
modulation. Satellites up to the order l=+3 and l=-4 can be resolved,
proving coherently grown superstructures in the growth direction. The
position of the satellite peaks give the superstucture periodicity from the
separation $\Delta(2\Theta)$ of the satellites of order l from the
fundamental Bragg peak \cite{art:zabel90}:

\begin{eqnarray}
\Lambda=\lambda/{[2\cdot l\Delta(\Theta)\cdot \cos(\Theta)]}
\end{eqnarray}

From this relation we get a superlattice period of 5.8 nm, in good
agreement with the value derived from the small angle x-ray reflectivity.
From the width of the satellite peaks at half maximum (FWHM)
$\Delta(2\Theta)$ we can derive the out-of-plane coherence length of the
superstructure $D_{coh}$ using the Scherrer equation

\begin{eqnarray}
D_{coh}=\lambda/[\Delta(2\Theta)\cdot \cos(\Theta)]
\end{eqnarray}

We estimate $D_{coh}=60$ nm i.e. comprising about 10 superlattice
periods. The fundamental Bragg peak in {\bf{Fig.6b}} is positioned at
$2\Theta=40.5^\circ$ i.e. at the middle position between the Au
(111)-Bragg peak at $2\Theta=38.5^\circ$ and the {\CuHeu} (220) peak at
$2\Theta=42.5^\circ$, as expected for a coherently strained superlattice.
Multilayers of similar hight quality can also be grown combining the
Heusler compounds {\CoGeHeu} and {\CoSnHeu} with Au.\\
In {\bf Table 2} we summarize the important parameters characterizing
the different multilayers with the Heusler compounds we have grown
successfully until now. As revealed by in-plane rocking scans all samples
exhibit a broad distribution of Bragg peaks in-plane and thus in are
polycrystalline  multilayers rather than superlattices.\\
Multilayers combining the Co-based Heusler alloys with V-interlayers can
also be grown. They possess sharp interfaces, however the out-of-plane
crystalline order is definitely worse than that we have obtained for the
multilayers with Au (see Table 2). We also have grown multilayers combining
two different Heusler phases. In {\bf Fig.7a} we present the  small angle
reflectivity scan and the large angle Bragg scan of the {\CuHeuCoGeHeu}
multilayer as an example. In the reflectivity one finds sharp superstructure
peaks up to the 4th order indicating a good quality of the layered structure
with sharp interfaces. From a fit of the reflectivity curve we determined a
superlattice periodicity $\Lambda=6.3$ nm. The Bragg scan close to the
(220)-peak exhibits one fundamental superlattice reflection at
$2\Theta=43.2^\circ$ and two weak satellite peaks giving a superlattice
periodicity of 6.4 nm. From the FWHM of the satellite peaks we estimate an
out-of-plane structural coherence length $D_{coh}$ of about 20 nm thus the
superstructure in the growth direction is coherent over about 3 periods. \\

Coming to the magnetic characterization of the multilayers, we have measured
the ferromagnetic saturation magnetization at 5 K for all multilayers and
summarized the results in Table 2. As discussed above, deviations of $M_s$
from the ideal bulk value can be taken as an indication of site disorder of
the Heusler alloys. By comparison with the bulk value of the magnetization
(see Table 1) one finds that most of the $M_s$ values of the multilayers are
definitely below  the bulk $M_{so} $. This partly can be attributed to the
lower preparation temperature of the multilayers (see Fig.5). For the
{\CuHeuAu} multilayer the magnetization is only about 12\% of the bulk
value, consistent with a strongly disordered B2-type of structure and spin
glass magnetic order. The {\CuHeuCoGeHeu} and the {\CuHeuCoSnHeu}
multilayers in Table 2 have a relative magnetization value $M_{s}/M_{0}>1$,
where one should note that $M_{0}$ refers to the saturation magnetization of
the Co-Heusler alloy alone. This clearly shows that the {\CuHeu}-layers in
the multilayers posses a substantial ferromagnetic magnetization definitely
larger than that observed for the single \CuHeu thin film prepared at the
same temperature. The reduced values of the saturation magnetization for
Co-based Heusler multilayers in combinatin with V and Au in Table 2 suggests
an intermixing at the
interfaces or an increased degree of site disorder.\\
More detailed insight into the metallurgical state and magnetism at the
interfaces can be gained by varying the thickness of the Heusler layers in
the multilayers. {\bf Fig.8} shows how the magnetic saturation magnetization
in the multilayers changes when decreasing the thickness of the Heusler
layers while keeping the thickness of the non magnetic layers constant at
3~nm. The ferromagnetic saturation magnetization breaks down for a thickness
of typically 1.5~nm in all systems. This result suggests that at the
interfaces of the multilayers there exists an intermediate layer of about
0.7~nm thickness which is metallurgically strongly disordered and not
ferromagnetic. We have recently shown in a separate investigation
\cite{Westerholt} that the interfaces in [\CoGeHeu/Au] multilayers develop
spin glass order at low temperatures leading to ferromagnetic hysteresis
loops with an unidirectional exchange anisotropy (so called exchange bias
effect). This result gives clear evidence for the existence of non
ferromagnetic interfaces. Quantitatively the decrease of the saturation
magnetization depicted in Fig.8 depends on the combination of both metals,
the multilayer [\CoSnHeu/V] developing the highest magnetization values in
the thickness range above 2~nm. In comparison the multilayer [\CoGeHeu/Au]
has a rather low value of the saturation magnetization in this thickness
range.

\section{\label{sec:level1}Summary and conclusions \protect\\}

In summary, we have shown that the Heusler phase {\CuHeu} can be
grown with high structural quality directly on  MgO (100) and
sapphire a-plane. For the half metallic Co-based Heusler compounds
{\CoSiHeu}, {\CoGeHeu} and {\CoSnHeu} it is possible to grow thin
films with flat surfaces, pure out-of-plane (110) texture and the
desired ordered $L2_1$ structure by using metallic seedlayers.
Principally this makes these compounds possible candidates
for applications in spin transport devices. \\
A crucial step in this direction is the test of the Co-based Heusler
compounds in the limit of very thin films and in combination with other
metallic layers. We have shown that for several combinations of the Heusler
compounds and nonmagnetic metals high quality, coherent multilayers can be
grown down to a thickness range of 1 nm for the Heusler phase. The magnetic
measurements however reveal that, depending of the specific combination of
materials and the thickness of the Heusler alloy layers, the saturation
magnetization is strongly lowered compared to the bulk value. Eventually,
for a thickness below typically about 1.5 nm, the Heusler layers are no
longer ferromagnetic. This result indicates that typically several
monolayers of the Heusler compounds at the interfaces are not ferromagnetic,
probably caused by alloying and (or) strong site disorder. This is not
unexpected, since an alloying at the interfaces can hardly be avoided in
real thin film systems and the chemical conditions for the phase formations
of a ternary compound at the interfaces are complex and virtually unknown.\\

We finally come to the question concerning the potential of the Co-based
Heusler alloys in the field of magnetoelectronics in the light of the
results presented here. The main problem will be to preserve the full spin
polarization predicted for the perfectly ordered Heusler structure in very
thin layers of real devices. We have shown that for the preparation
temperatures allowed in thin film heterostructures the formation of site
disorder cannot be completely avoided in the Co-based Heusler alloys. Site
disorder in the interior of the Heusler film is a critical factor, since it
must be expected that the full spin polarization is lost in disordered
Heusler alloys \cite{art:ishida2-98}. The question, to what extend some site
disorder is tolerable, i.e. leaves at least a high value for the spin
polarization, cannot be answered quantitatively at the moment, since
corresponding band structure calculations have not been published yet. The
existence of non ferromagnetic interfaces, which seem to be present in all
combinations of the Co-based Heusler alloys and other metals which we have
studied until now, causes a second problem, which might be even more
detrimental for the performance of spin transport devices. Necessarily the
spin polarization at the Fermi level will completely vanish in a non
ferromagnetic interlayer. Since in GMR with the current in the plane
(cip-geometry) the spin dependent electron scattering at the interfaces is
dominating \cite{art:Dieny}, non ferromagnetic interfaces are expected to
reduce the GMR-value strongly. This is in accord with our first results of
magnetoresistance measurements for spin valve systems using the Co-based
Heusler alloys, which reveal very small values for the GMR effect
\cite{Bergmann}. Possibly one can overcome this problem by using the GMR
geometry with the current perpendicular to the plane (cpp-geometrty). In
this geometry one can use much larger thicknesses of the ferromagnetic
layers and the spin asymmetry of the electron scattering in the interior of
the ferromagnetic layers gives an important contribution to the GMR
\cite{art:Gijs}. However, concerning technical applications the cpp-geometry
seems not very useful. The alternative choice would be to search for other
material combinations or preparation methods with less interdiffusion at the
interfaces.

\begin{ack}
The authors thank the DFG for financial support of this work
within the SFB 491, P. Stauche for the preparation of the alloy
targets and S. Erdt-B\"{o}hm for running the sputtering equipment.
\end{ack}




\section{\label{sec:level1}Figure Captions\protect\\ }

{\bf Fig.1}\\
Out-of-plane x-ray Bragg-scan of a {\CuHeu} film on MgO(100). The inset
shows the in-plane rocking scan of the Heusler (200) peak.\\

{\bf Fig.2}\\
(a)Out-of-plane Bragg-scan of a {\CuHeu} film grown directly on
 Al$_{2}$O$_{3}$ and (b) low angle x-ray reflectivity spectrum of the same
    film.\\

 {\bf Fig.3}\\
Out-of-plane Bragg-scan of a {\CoSnHeu} film grown on an Au seed layer.\\

 {\bf Fig.4}\\
 Small angle x-ray reflectivity scan or a {\CoGeHeu} film on Al$_{2}$O$_{3}$ with
 a V-seedlayer\\

{\bf Fig.5}\\
Saturation magnetization of {\CoGeHeu}, {\CoSnHeu} and {\CuHeu} versus
the substrate temperature during preparation.\\

 {\bf Fig.6}\\
(a) Small angle x-ray reflectivity scan of a multilayer {\CuHeuAu} with a
nominal thickness of 3 nm for Au and {\CuHeu} and (b) out-of-plane Bragg
scan of the same multilayer. The numbers in the figure denote the order
of the superlattice reflections and the order of the sattelites\\

  {\bf Fig.7}\\
(a) Small angle reflectivity scan of a {\CuHeuCoGeHeu} multilayer and (b)
large angle Bragg-scan for the same sample.\\

  {\bf Fig.8}\\
Ferromagnetic saturation magnetization of multilayers measured at 5 K as
a function of the thickness of the Co-Heusler layers. The thickness of
the  other layer is kept constant at 3 nm.\\

\vspace*{\fill}
\pagebreak[4]

\begin{table}[]
    \centering
    \scriptsize
\begin{tabular}{|c|c|c|c|c|c|c|c|c|}
  \hline
  \textbf{Phase} & \textbf{Prep.}& \textbf{Substr./seed layer} & \textbf{Texture} & {\textbf{Rocking}} & \multicolumn{2}{|c|}{{\textbf{Lattice parameter}}} & \multicolumn{2}{|c|}{\textbf{Magnet-}} \\
                & \textbf{temp.}& \textbf{} &  & \textbf{width(220)}- & \multicolumn{2}{|c|}{\textbf{(nm)}} & \multicolumn{2}{|c|}{\textbf{ization}} \\
                & \textbf{({\deg}C)} & &  & \textbf{peak ({\deg}}) &  \multicolumn{2}{|c|}{}& \multicolumn{2}{|c|}{\textbf{(emu/g)}} \\ 
                & & &  & & \textbf{bulk} & \textbf{film} & \textbf{bulk} & \textbf{film} \\
                \hline
  Cu$_{2}$MnAl & 470 & Al$_{2}$O$_{3}$ a-plane & (110) & 0.8 & 0.5962 & 0.5907 & 98 & 62 \\
  Cu$_{2}$MnAl & 470 & MgO (100) & (100) & 0.16 &  & 0.5922 & & 40 \\
  \hline
  Co$_{2}$MnSi & 470 & Al$_{2}$O$_{3}$ a-plane/Cr & (110) & 4 & 0.5654 & 0.5688 & 138 & 98 \\ 
  Co$_{2}$MnSi & 470 & MgO (100)/Cr& (100) & 10 &  & 0.5670 &  & 100 \\ 
  \hline
  Co$_{2}$MnGe & 470 & Al$_{2}$O$_{3}$ a-plane/V & (110) & 3 & 0.5743 & 0.5766 & 111 & 103 \\ 
  Co$_{2}$MnGe & 470 & MgO (100)/V & (100) & 5 & & 0.5803 & & 107 \\
   \hline
  Co$_{2}$MnSn & 470 & Al$_{2}$O$_{3}$ a-plane/Au & (110) & 3 & 0.6000 & 0.6003 & 91 & 87 \\ 
  Co$_{2}$MnSn &470& MgO (100)/Au& (100) & 6 & & 0.6011 & & 80 \\
  \hline
\end{tabular}
        \caption{Structural parameters and saturation magnetization measured at 5 K for the Heusler films
        grown  on Al$_{2}$O$_{3}$ a-plane or MgO (100)
        \label{tab1}}
\end{table}

\begin{table}[ht]
    \centering
    \scriptsize
\begin{tabular}{|c|c|c|c|c|c|c|c|c|}
  \hline
  \textbf{Multilayer} &  \textbf{Texture} & \textbf{Period} & \textbf{Coherence} & \textbf{Lattice-} & {\textbf{$M_{s}/M_{0}$}} \\
                &    & \textbf{length(nm)} & \textbf{length(nm)} &  \textbf{parameter  (nm)} & \\
                &    &                     &                     &\textbf{out-of-plane} & \\
                 \hline
  {\CuHeuAu} &     (110)/(111) & 5.7 & 60 & 0.610 / 0.400 & 0.12 \\ \hline
  {\CuHeuCoSnHeu}  & (110)     & 6.0 & 20 & 0.598 / 0.598 & 1.45 \\ \hline
  {\CuHeuCoGeHeu}  & (110)     & 6.4 & 18 & 0.588 / 0.588 & 1.09 \\ \hline
  {\CoSnHeuAu} &  (110)/(111)  & 5.9 & 50 & 0.615 / 0.400 & 0.68 \\ \hline
  {\CoSnHeuV} &      (110)     & 6.3 & 30 & 0.596 / 0.299 & 0.71 \\ \hline
  {\CoGeHeuV} &      (110)     & 6.2 & 35 & 0.584 / 0.292 & 0.70 \\ \hline
  {\CoGeHeuAu} &  (110)/(111)  & 5.9 & 70 & 0.620 / 0.409 & 0.47 \\ \hline
\end{tabular}
        \caption{Structural parameters of the Heusler multilayers grown
        on Al$_{2}$O$_{3}$ a-plane and the relative saturation magnetization
        measured at 5 K. $M_{0}$ denotes the saturation magnetization of
        the bulk Co-Heusler compounds (see Table 1)}
\end{table}

\begin{figure}[ht]
    \centering
       \includegraphics[clip=true,keepaspectratio=true,width=0.7\linewidth]{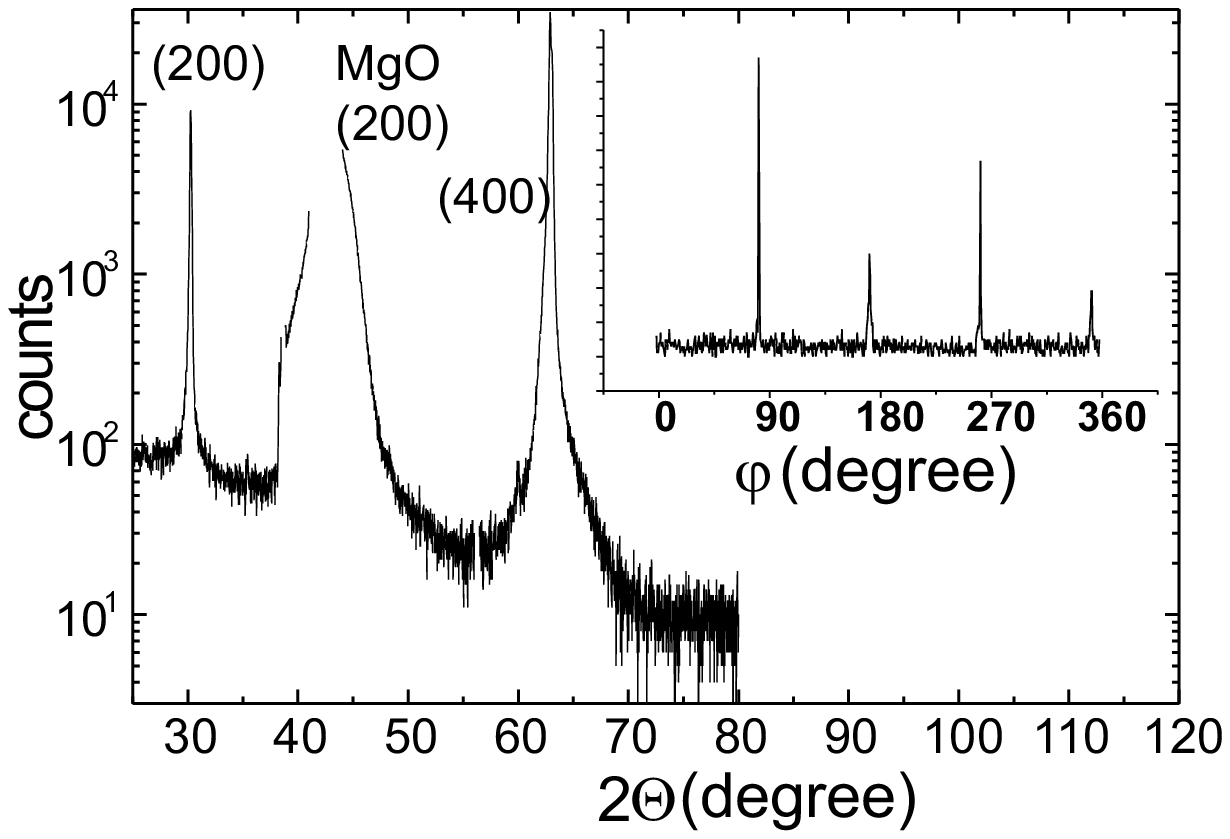}
    \caption{\label{Fig1}}
\end{figure}

\begin{figure}[ht]
    \begin{center}
    \subfigure[\label{Fig2a}]
    {\includegraphics[clip=true,keepaspectratio=true,width=0.7\linewidth]{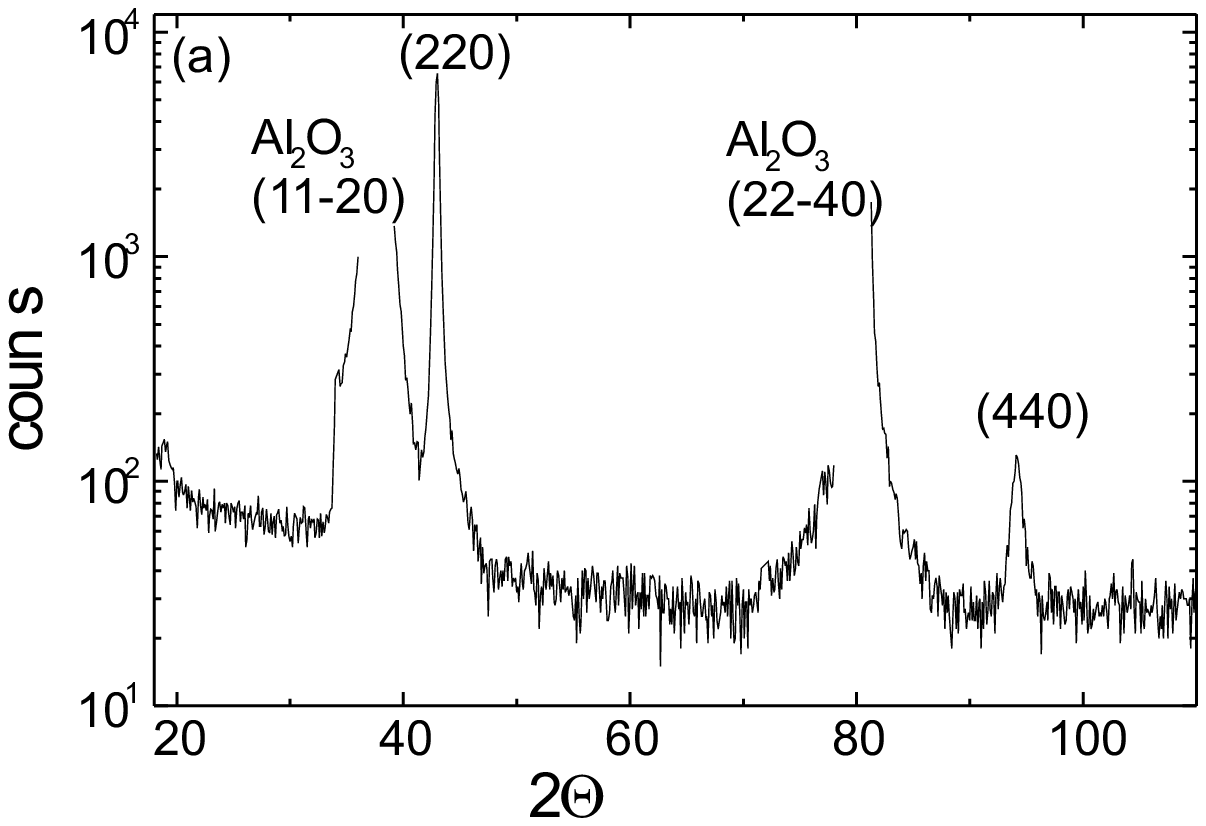}}
    \hfill
    \subfigure[\label{Fig2b}]
    {\includegraphics[clip=true,keepaspectratio=true,width=0.7\linewidth]{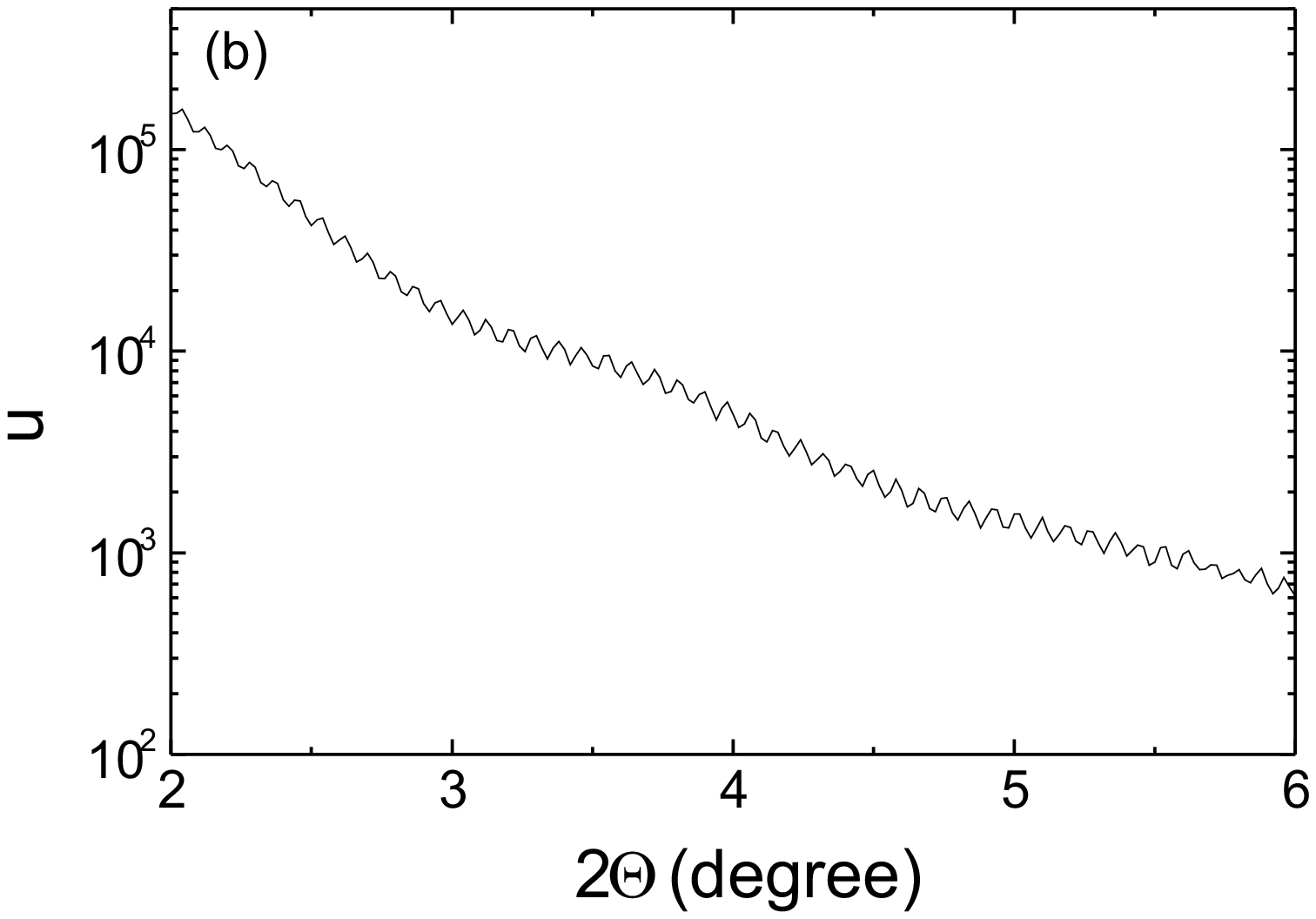}}
   \end{center}
        \caption{\label{Fig2 tot.}}
\end{figure}

\begin{figure}[ht]
    \centering
       \includegraphics[clip=true,keepaspectratio=true,width=0.7\linewidth]{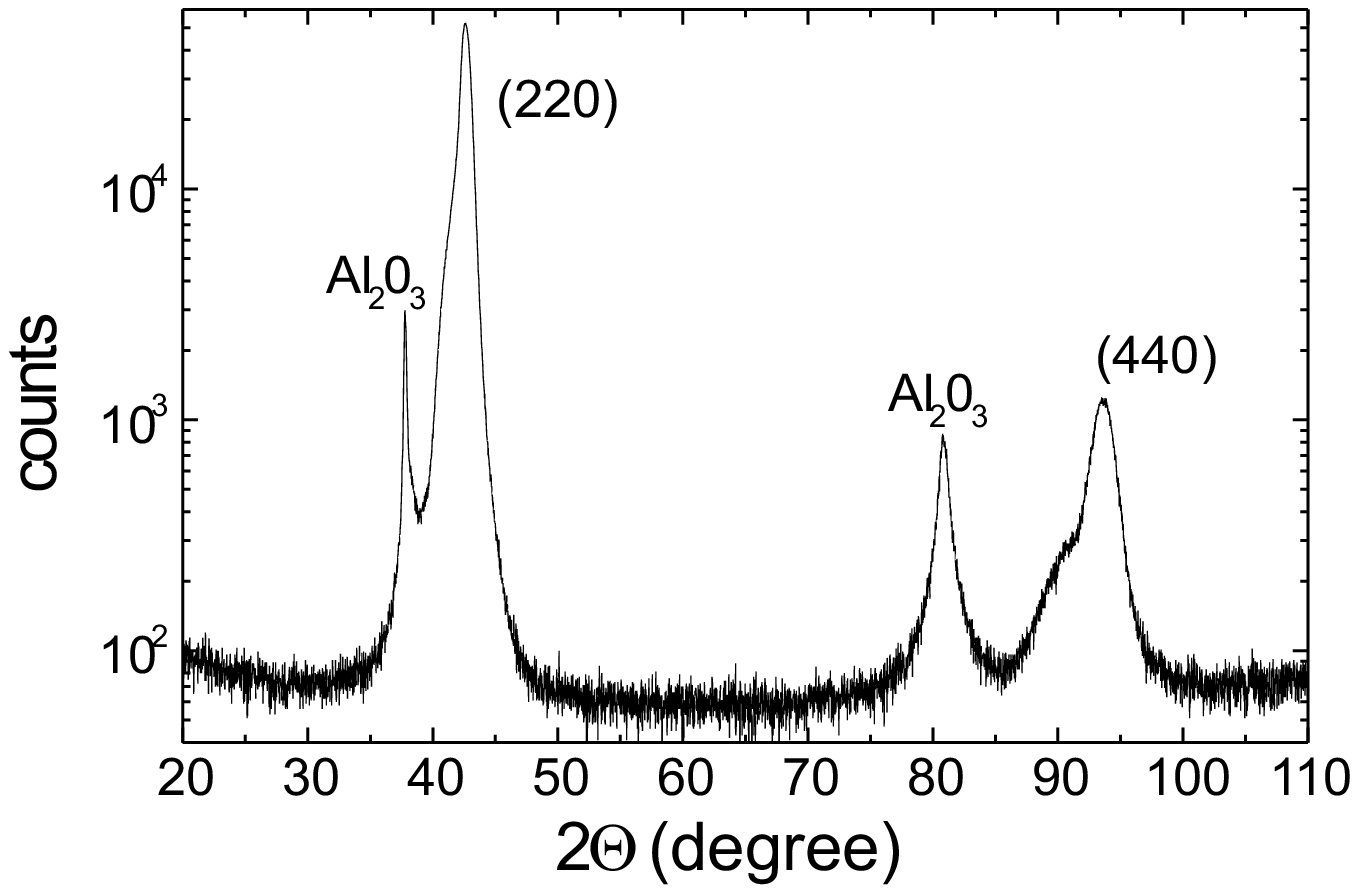}
    \caption{ \label{Fig3}}

\end{figure}

\begin{figure}[]
    \centering
       \includegraphics[clip=true,keepaspectratio=true,width=0.7\linewidth]{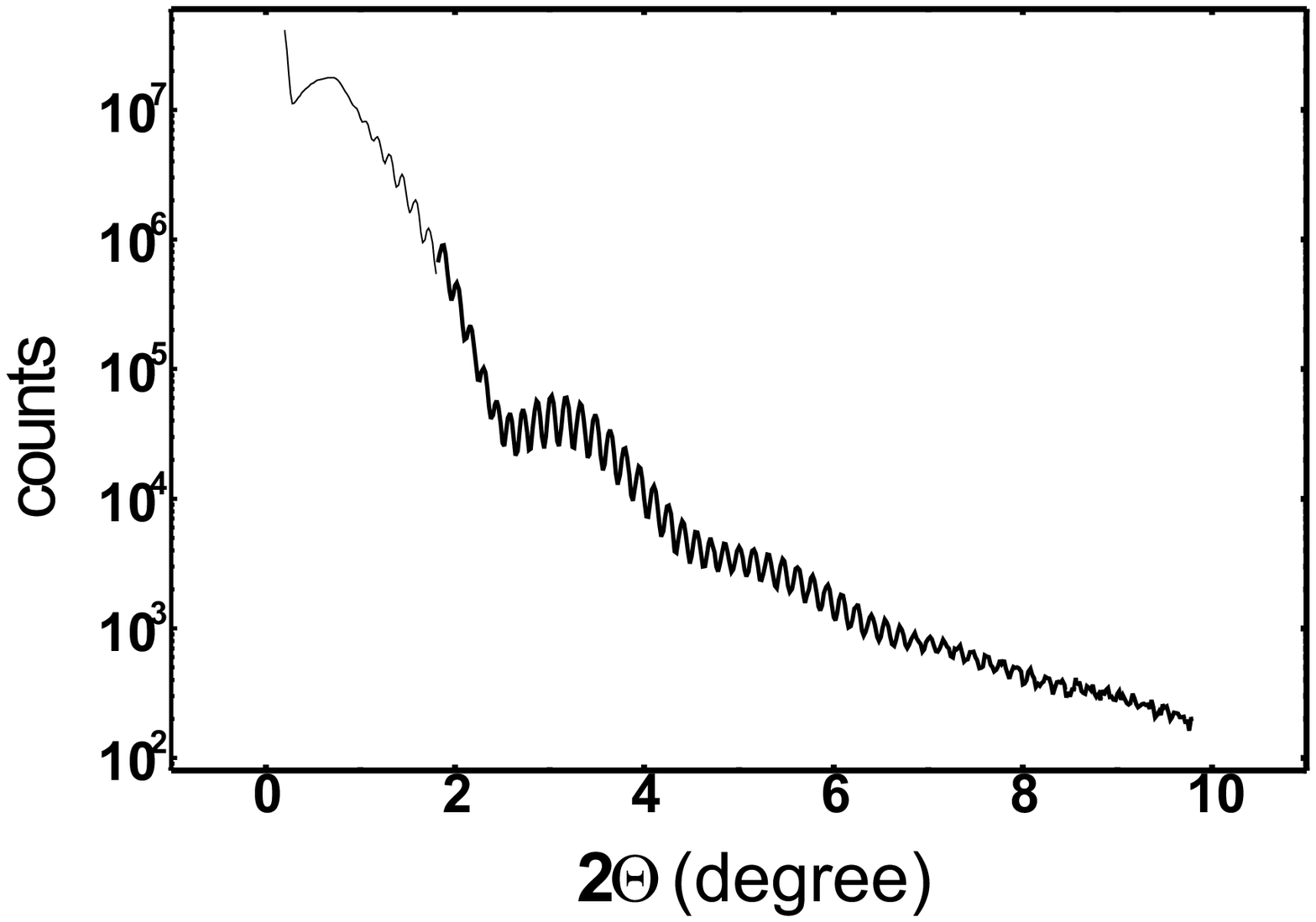}
    \caption{ \label{Fig4}}
\end{figure}

\begin{figure}[]
    \centering
       \includegraphics[clip=true,keepaspectratio=true,width=0.7\linewidth]{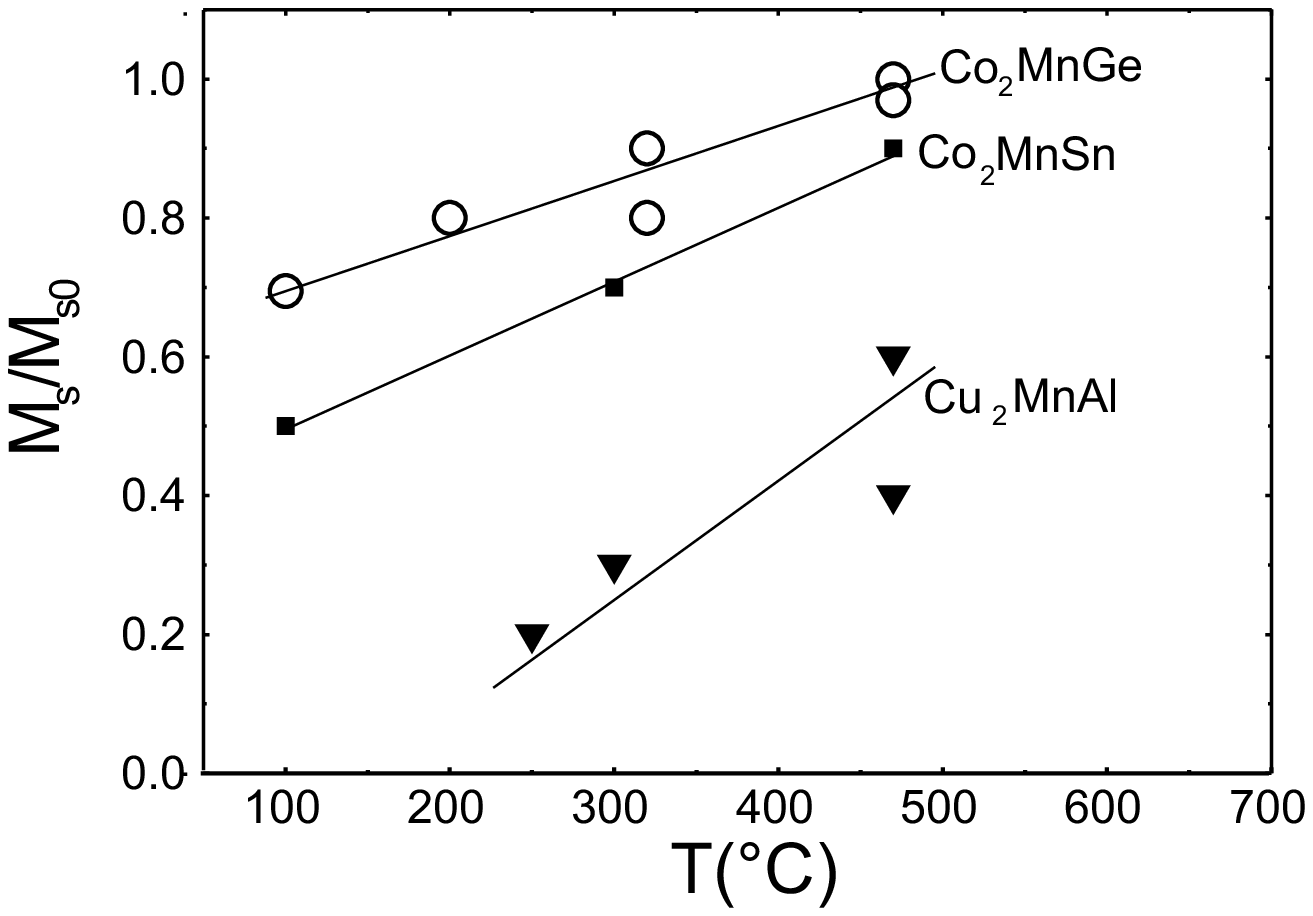}
    \caption{ \label{Fig5}}
\end{figure}

\begin{figure}[]
    \begin{center}
    \subfigure[\label{Fig6a}]
    {\includegraphics[clip=true,keepaspectratio=true,width=0.7\linewidth]{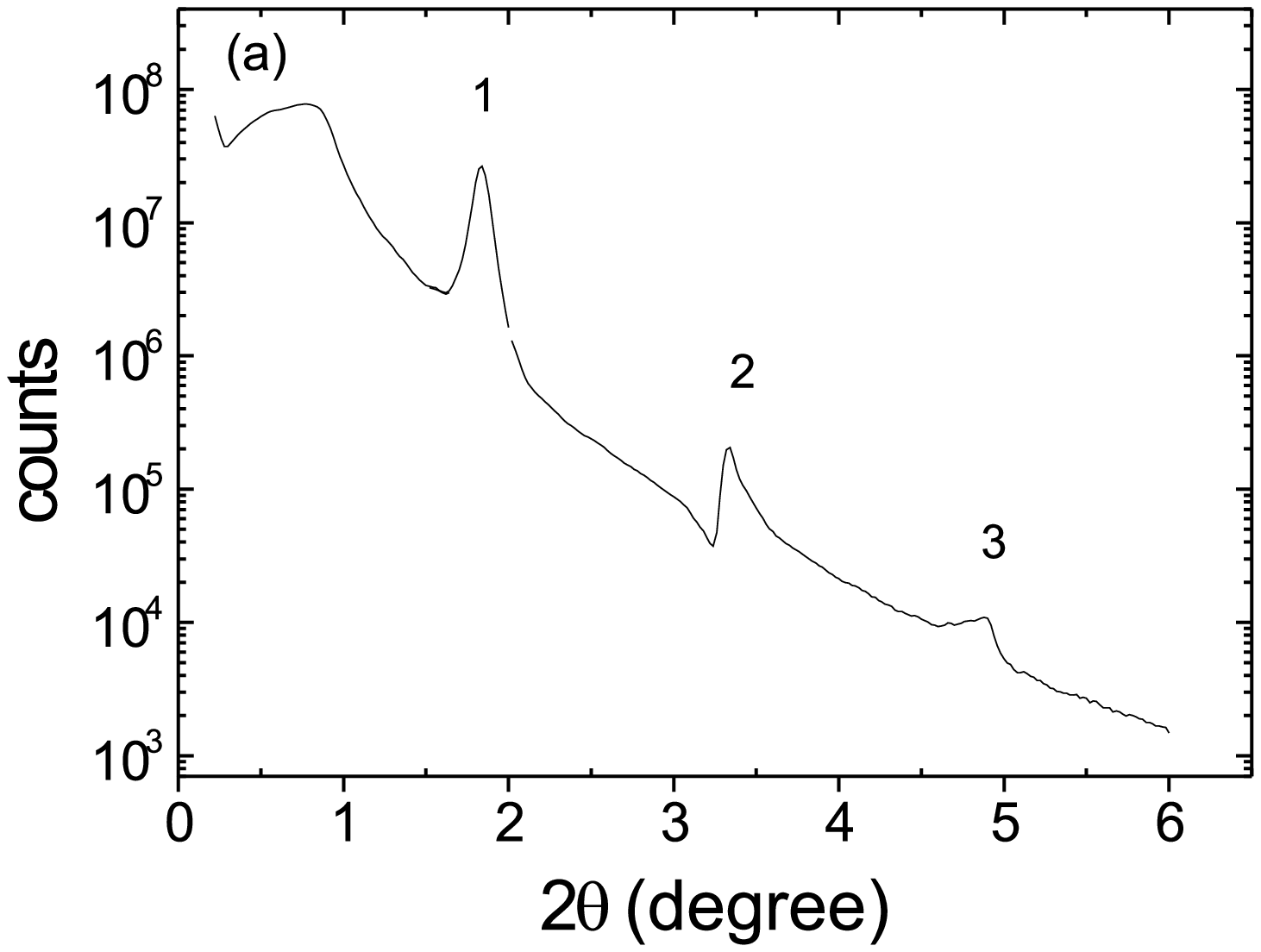}}
    \hfill
    \subfigure[\label{Fig6b}]
    {\includegraphics[clip=true,keepaspectratio=true,width=0.7\linewidth]{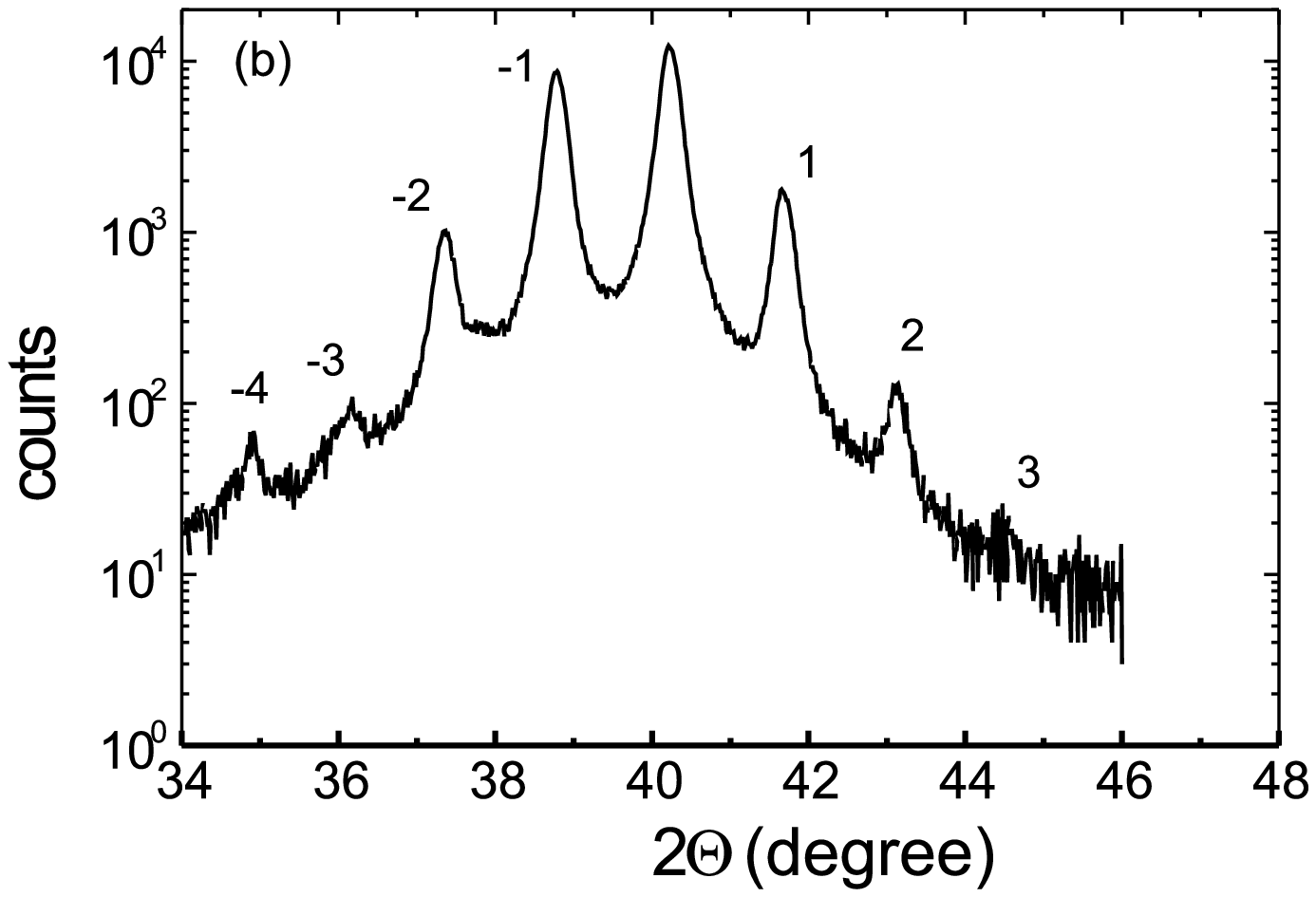}}
   \end{center}
        \caption{\label{Fig6_tot}}
\end{figure}

\begin{figure}[ht]
    \begin{center}
    \subfigure[\label{Fig7a}]
    {\includegraphics[clip=true,keepaspectratio=true,width=0.7\linewidth]{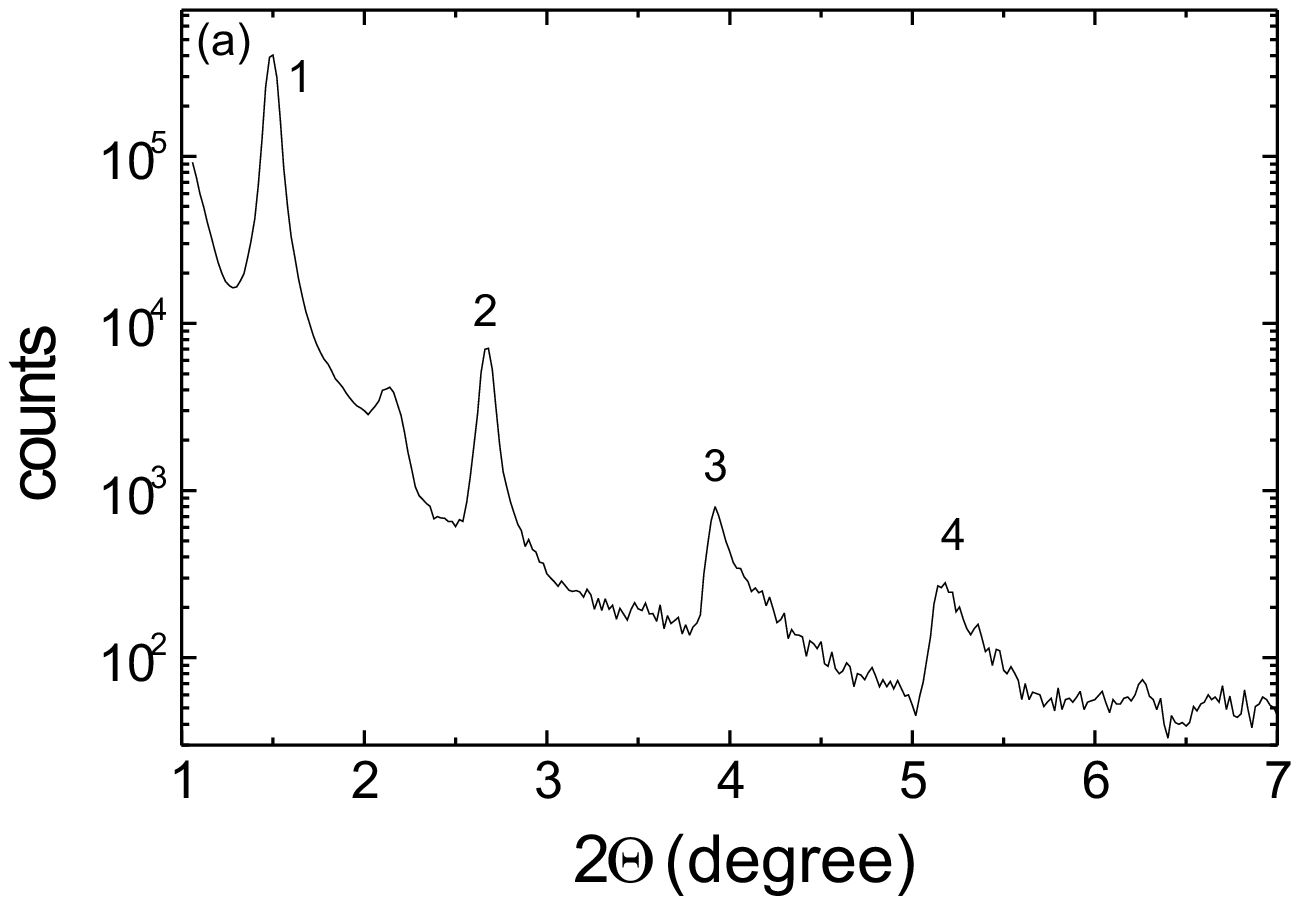}}
    \hfill
    \subfigure[\label{Fig7b}]
    {\includegraphics[clip=true,keepaspectratio=true,width=0.7\linewidth]{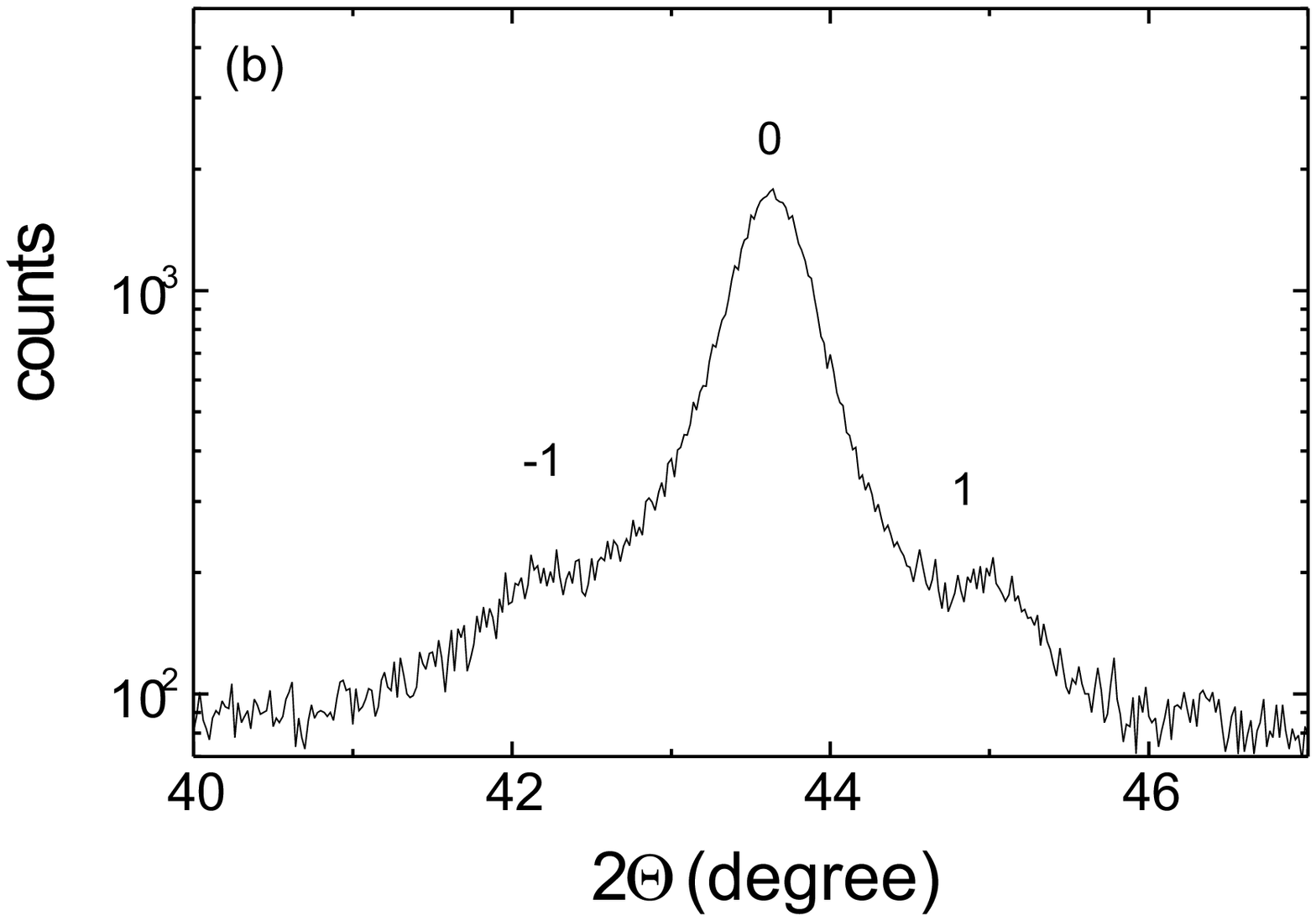}}
   \end{center}
        \caption{\label{Fig.7_tot}}
\end{figure}

\begin{figure}[ht]
    \centering
       {\includegraphics[clip=true,keepaspectratio=true,width=0.7\linewidth]{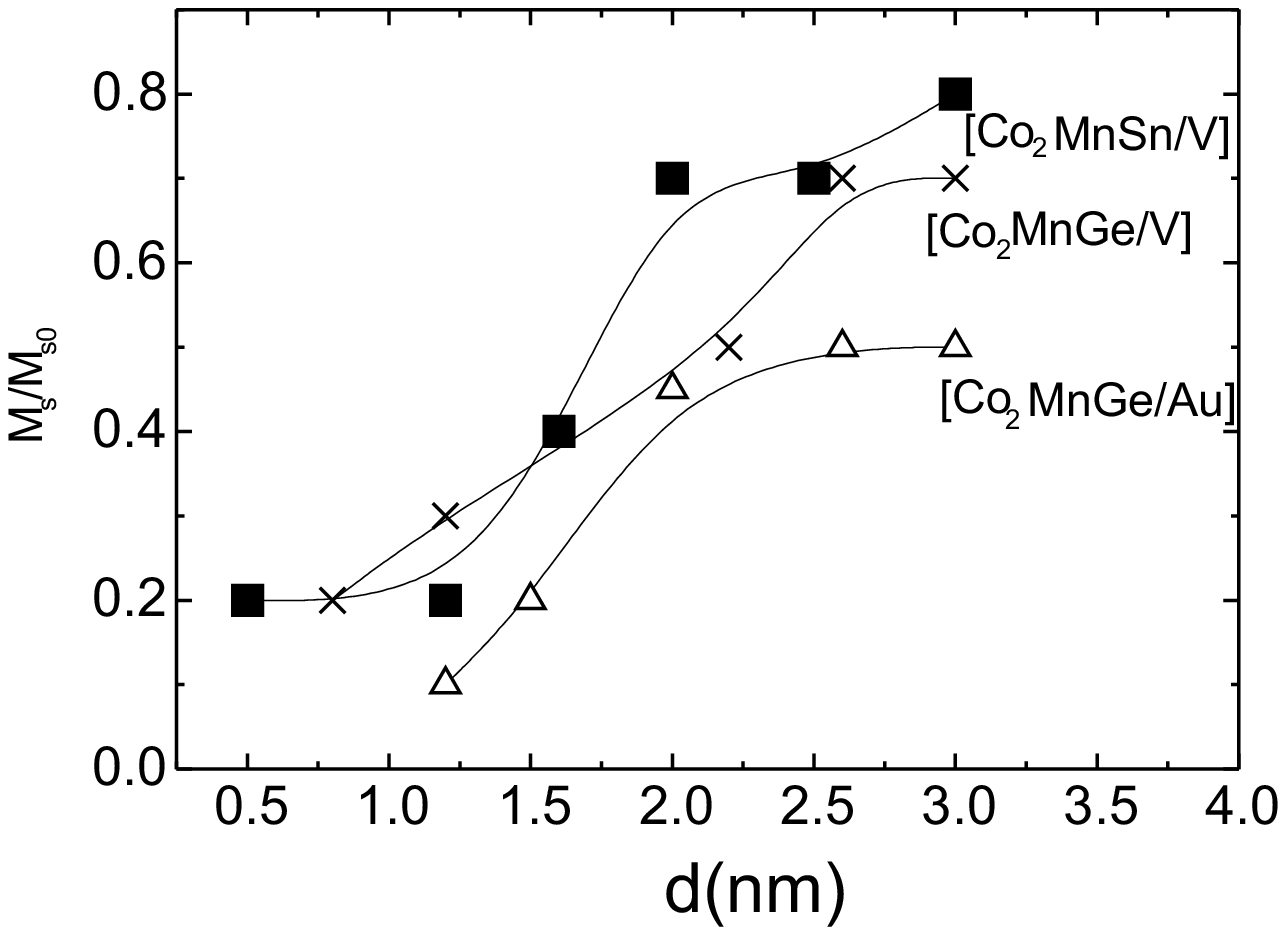}}
    \caption{ \label{Fig8}}
\end{figure}


\clearpage

\begin{thebibliography}{99}



\bibitem{art:park98}
J.~Park, E.~Vescovo, E.~Kim, C.~Kwon, R.~Ramesh, T.~Venkatesan, Nature 392
  (1998) 794.

\bibitem{art:Moodera}
J.~Moodera, R.~Kinder, L, T.~Wong, R.~Meservey, Phys. Rev. Lett. 74 (1998)
  3273.

\bibitem{art:Dieny}
B.~Dieny, J. Magn. Magn. Mat. 136 (1994) 335.

\bibitem{art:Schwarz}
K.~Schwarz, J. Phys. F: Met. Phys. 16 (1986) 7934.

\bibitem{pbk:ziebeck}
Ziebeck, Webster, Landolt-B\"{o}rnstein New Series III/19c, Springer-Verlag,
  1988.

\bibitem{art:deGroot83}
R.~de~Groot, P.~van Engen, Phys. Rev. Letters 50 (1983) 2024.

\bibitem{art:ishida2-98}
S.~Ishida, T.~Masaki, S.~Fujii, S.~Asano, Physica B 245 (1998) 1.

\bibitem{art:ishida1-95}
S.~Ishida, S.~Fujii, S.~Kashiwagi, Journal of the Physical Society of Japan
64
  (1995) 2152.

\bibitem{art:kabani90}
R.~Kabani, M.~Terada, A.~Roshko, J.~Moodera, J. Appl. Phys. 67 (1990) 4898.

\bibitem{art:caballero99}
J.~Caballero, A.~Reilly, Y.~Hao, J.~Bass, W.~Pratt, F.~Petroff,
J.~Childress,
  J. Magn. Magn. Mat. 55 (1999) 198.

\bibitem{art:Kautzky}
M.~Kautzky, F.~Mancoff, J.~Bobo, P.~Johnson, R.~White, B.~Clemens, J. Appl.
  Phys. 81 (1997) 4026.


\bibitem{art:Orgassa}
D.~Orgassa, H.~Fujiwara, T.~C.~Schulthess, W.~H.~Butler,
  Phys. Rev B 60 (1999) 13237



\bibitem{art:raphael01}
M.~Raphael, B.~Ravel, M.~Willard, S.~Cheng, D.~Das, R.~Stout, K.~Bussmann,
  Appl. Phys. Lett. 79 (2001) 4396.

\bibitem{art:ambrose01}
T.~Ambrose, J.~Krebs, G.~Prinz, J. Appl. Phys. 89 (2001) 7522.

\bibitem{art:Geiersbach}
U.~Geiersbach, A.~Bergmann, K.~Westerholt, J. Magn. Magn. Mat. 240 (2002)
546.

\bibitem{art:kuebler83}
J.~K\"{u}bler, A.~Williams, C.~Sommers, Phys. Rev. B 28 (1983) 1745.

\bibitem{art:Mancoff}
F.~B. Mancoff, J.~Bobo, O.~Richter, J. Mater. Res. 14 (1999) 1560.

\bibitem{Bruno}
P.~Bruno, Europhys. Lett. (1993) 615

\bibitem{art:Soltys}
J.~Soltys, Acta. Phys. Pol. A 56 (1979) 227.

\bibitem{art:paratt54}
L.~Paratt, Phys. Rev. 95 (1954) 359.

\bibitem{art:Bacon}
G.~Bacon, J.~Plant, J. Phys. F: Metal Physics 1 (1971) 524.

\bibitem{art:Robinson}
J.~Robinson, P.~McCormick, R.~Street, J. Phys. Condens. Matt. 7 (1995) 4259.

\bibitem{art:Taylor}
R.~Taylor, C.~Tsuei, Sol. St. Comm. 41 (1982) 503.

\bibitem{art:Webster}
P.~Webster, J. Phys. Chem. Sol. 32 (1971) 1221.

\bibitem{art:zabel90}
H.~Zabel, Advances in Sol. St. Phys. 30 (1990) 197.

\bibitem{Westerholt}
K. Westerholt, U. Geiersbach and A.~Bergmann, appears in J. Magn. Magn. Mat.
(2002)

\bibitem{Bergmann}
 A.~Bergmann and K.~Westerholt, in  preparation


\bibitem{art:Gijs}
M.~Gijs, G.~Bauer, Advances in Physics 46 (1997) 286.

\end{thebibliography}
\end{document}